\documentclass[aps,pra,twocolumn,showpacs]{revtex4-1}

\newcommand{\ket}[1]{| \, #1 \rangle}

\newcommand{\bra}[1]{\langle #1 \, |}

\def\>{\rangle}
\def\<{\langle}

\usepackage{graphicx}
\usepackage{amsmath}
\usepackage{amsfonts,amssymb,graphicx,latexsym,color}
\usepackage{bm}

\begin{document}

\title{Coherent quantum effects through dispersive bosonic media}
\author{Sai-Yun Ye,$^{1, 2}$ Zhen-Biao Yang,$^{2}$ Shi-Biao Zheng,$^{2}$ and Alessio Serafini$^{1}$}
\affiliation{$^{1}$ Department of Physics and Astronomy, University College London, Gower Street, London WC1E 6BT, UK}
\affiliation{$^{2}$ Department of Physics, Fuzhou University, Fuzhou 350002, P.R.~China}

\pacs{03.67.Lx, 42.50.Ex, 42.50.Ct, 42.81.Qb}

\date{\today }

\begin{abstract}
The coherent evolution of two atomic qubits mediated by a set of bosonic field modes is investigated.
By assuming a specific encoding of the quantum states in the internal levels of the two atoms
we show that entangling quantum gates can be realised, with high fidelity, even when a large number of mediating modes is involved.
The effect of losses and imperfections on the gates' operation is also considered in detail.
\end{abstract}
\maketitle

\section{Introduction}

There is great emphasis, in current research, on the identification of physical systems and conditions
where coherent quantum effects are susceptible to emerge.
In fact, the achievements of quantum communication and key distribution \cite{gisin02},
quantum metrology \cite{giova04},
and the prospects opened by quantum computation \cite{book,ladd10}
are all centred on quantum
coherence. Also, the latter has been shown to occur in biological proteins and
is suspected to play a role in some mechanisms
of energy transfer in biological structures \cite{bio}.
In this context, it is extremely relevant, for both fundamental and applied reasons,
to single out systems and operating regimes where
coherent effects can take place, and to analyze them in detail.

A recurrent paradigm for quantum information processing
comprises a set of static qubits where the information is stored
(usually embodied by internal levels of atoms, ions or quantum dots in solid state devices),
interacting through a bosonic field (light or vibrational phonons).
Countless variations of this paradigm have been proposed
and studied, in both theory and practice, over the last fifteen years,
with peculiar advantages and disadvantages for the
realization of controlled coherent manipulations \cite{cirac95,cirac97,vanenk,1,13,17,wineland,
clark,bla,xiao,barrett,20,zhou,beige1,serafozzi,qubus,27,6,zou,28,beige2,26,8,zhenbiao,noiartri}.

However, in all of these studies, only one of the bosonic field modes
-- or a selected few, in more sophisticated settings -- are assumed to mediate the interaction
between the qubits. This assumption may be reasonable if the space where
the bosonic field modes resonate -- which we will henceforth refer to as the ``fibre'' -- is small,
so that the frequency spacing is very large, and if
the experimental control is very accurate.
Such conditions are not always met in practical situations:
often more than one mode mediates the interaction between the qubits,
for instance when the fibre is too long, or when unwanted degrees of freedom
outside direct experimental control enter the dynamics.
When such situations occur, one expects the capability of the system
to mediate the controlled, coherent evolutions for the two qubits to be
degraded, simply because the quantum information disperses among a
greater number of degrees of freedom. The question then arises of whether
controlled dynamics are still possible in ``dispersive'' bosonic media.

This paper intends to launch an investigation to understand whether, to what extent,
and under what conditions may
coherent quantum effects be mediated by multiple degrees of freedom.
Notice that such a question is relevant not only to quantum information processing
(where, it may be argued, the relative flexibility in the engineering of quantum systems
sometimes allows one to circumvent the effect of undesired degrees of freedom),
but also to other, more fundamental applications of quantum theory, like in the modeling of
certain light-harvesting biological compounds,
where several frequencies of the light field are thought to coherently mediate
the transfers of energy.

\begin{figure}[b!]
\includegraphics[scale=0.6]{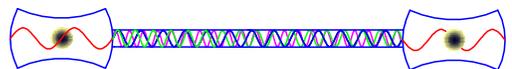}
\caption{(color online). The prototype: two remote nodes, where atoms are trapped in, are linked by a media containing many bosonic modes at different frequencies.} \label{fig1}
\end{figure}

In this study, we shall refer to a specific basic
`prototype' which has drawn considerable attention as
a candidate for `distributed' quantum information processing.
Two distant atoms are trapped in two remote optical micro-cavities, see Fig.~\ref{fig1},
and interact locally with one of the cavity modes.
The cavities are then in turn connected through an optical fibre where
{\em many} light modes couple to the cavity modes.
In practice, this prototype might represent two atoms or ions trapped in
micro-fabricated optical cavities directly coupled to the same
integrated optical fibre \cite{microcav}.
However, our results will apply to more general systems
sharing the same formal description.
In this respect notice that, by transforming to
normal modes of the cavities plus fibre field, our system can be shown to be
equivalent to two qubits whose interaction is directly mediated by a set of bosonic modes
(without the two interfacing cavities fields).

To demonstrate that controlled coherent quantum evolutions are possible, we shall focus
on the realisation of a quantum controlled-z ({\sc CZ}) gate for the two qubits. The perfect implementation
of such an operation would allow for maximal entanglement to be generated and,
supplemented with single-qubit gates,
would provide one with a universal set of quantum gates,
on any number of qubits \cite{bremner}.
No specific initial state will be assumed: the fidelity of the gate will be averaged over the
unitarily-invariant Haar measure of the two-qubit state space.
By adopting a particular structure for the atomic levels and qubit encoding,
we will show that, in several situations, this non-trivial controlled quantum dynamics
of the two qubits
can be approximated with high fidelity, even if
more than one mode mediates the interaction between the distant systems.
This also implies that the creation of nearly maximal entanglement between
the two qubits is possible.

\section{Methods}

The main obstacles to our programme are twofold. On the one hand, we have to deal with the
customary computational difficulty of describing a many-body quantum system.
On the other hand, we want to identify a situation where a controlled entangling quantum gate
can be mediated by several degrees of freedom, so that we have to find a way to
reduce the effects of dispersion through such a medium.

\begin{figure}[t!]
\includegraphics[width=3in]{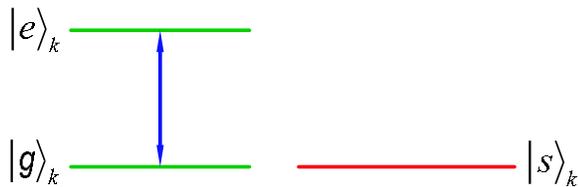}
\caption{(color online). The atomic level configuration of the two atoms ($k=1,2$).
Notice that level $\ket{s}_1$ does not play any role in our scheme.
Level $\ket{e}_2$, instead, enters the dynamics but is not used to encode
any quantum information.} \label{fig2}
\end{figure}

Both these issues will be solved by
encoding the two qubits asymmetrically -- {\em i.e.}, differently in the two sites --
in one excited state and two kinds of ground states,
following a strategy introduced in Ref.~\cite{zhenbiao}.
In the first cavity, labelled by $1$,
the qubit is encoded in the ground state $\ket{g}_1$ and in the excited state $\ket{e}_1$,
which are coupled to each other via the cavity mode, by a rotating-wave Hamiltonian.
In the second cavity, labelled by $2$,
the ground state $\ket{g}_2$ is also coupled to the excited state $\ket{e}_2$, but
the qubit is encoded in the ground states $\ket{g}_2$ and $\ket{s}_2$,
and the latter is not coupled to any state and does not evolve at all in
the dynamics we consider.
This configuration allows us to study the realisation of the {\sc CZ} gate by
restricting to the single excitation subspace, thus both reducing the computational effort required
-- if $N$ is the number of mediating modes, the size of the relevant Hilbert subspace
only scales like $N$ -- and limiting the dispersive effect of the modes.

We shall assume
the cavity fields $a_1$ and $a_2$ to be at resonance
with the atomic transitions, so that the rotating-wave approximation holds,
and shall have them interacting with $N$ modes of the ``fibre'' at different frequencies.
In the frame rotating
at the cavity frequency, the total Hamiltonian of the system is then
\begin{widetext}
\begin{equation}
\hat{H}=\sum\limits_{j=1}^N\Delta w_jb_j^{\dag}b_j
+ \left(\sum\limits_{k=1}^{2}g_ka_k\mid
e\rangle _k\langle g\mid
_k + h.c. \right)
+\left(\sum\limits_{k=1,j=1}^{k=2,j=N} v _k b_j a_k^{\dag}+h.c.\right),
\end{equation}
where $\Delta w_j$ is the frequency difference between the $k$th mediating
mode and the cavity mode, $a_k^{\dag}$ ($a_k$) and $b_j^{\dag}$($b_j$) are the
creation (annihilation) operators for the cavity modes and the fibre modes,
respectively, $ v  _k$ denotes the coupling strength between the mode of cavity $k$
and the mediating modes (possibly including a phase),
and $g_k$ represents the coupling strength between atom and field in cavity $k$.
Note that the fibre-cavities couplings depend on the cavity but are assumed to be the same for all the
fibre modes. The emphasis in our analysis is rather on the effect of the detunings $\Delta w_j$
of the fibre modes.

Also, we will consider both losses of the cavities and fibre and spontaneous emission from the atoms
(by far the main sources of decoherence at optical frequencies), so that
our dynamics is governed by the master equation
\begin{equation}
\dot{\rho }=-i[\hat{H},\rho ]+\kappa \sum\limits_{k=1}^2L[a_k]\rho
+\gamma \sum\limits_{k=1}^2L[\sigma ^{-}]\rho +\Gamma
\sum\limits_{j=1}^NL[b_k]\rho \; ,  \label{master}
\end{equation}
where $L[\hat o]=2\hat o\rho \hat o^{\dag}-\hat o^{\dag}\hat o\rho -\rho \hat o^{\dag}%
\hat o$ is for operator $\hat o$, and $\gamma$, $\kappa$ and $\Gamma$ are the atomic
spontaneous emission rate and the loss rates of the cavities -- assumed, for simplicity, to be identical --
and of the fibre, respectively.

\begin{figure}[t!]
\includegraphics[width=18cm,height=4.9cm]{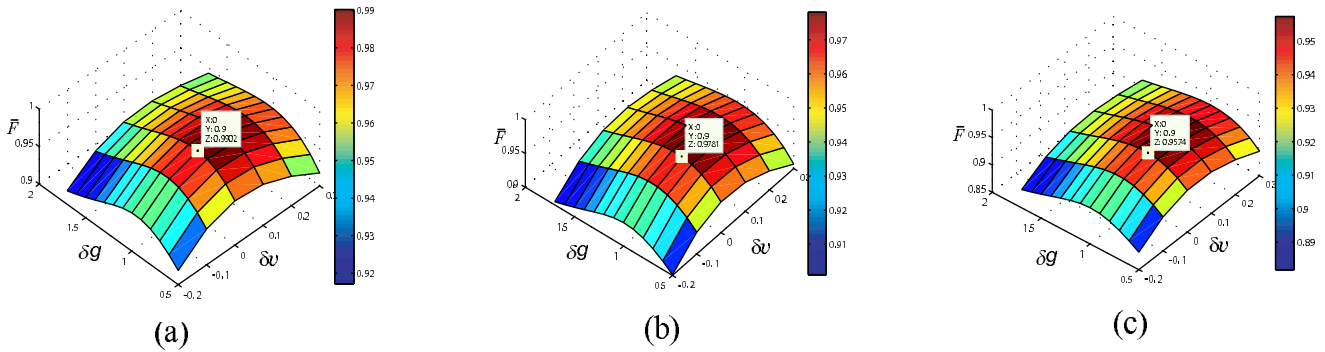}
\caption{(color online). The average gate fidelity $\overline{F}$ for two detuned modes versus $\delta g$ and
$\delta v$ not taking losses into account, for $\Delta=0.1g$ and $gt=4.6$ (a),
$\Delta=0.2g$ and $gt=4.56$ (b), $\Delta=0.3g$ and $gt=4.54$ (c).} \label{fig3}
\end{figure}

Initially, all the field modes including the cavity and
mediating modes are in the vacuum state $\ket{0}_f$.
Moreover, to check the reliability of
the controlled phase gate, we consider the generic input state of two qubits
\begin{equation}
\mid \Psi \rangle _{in}=\alpha \mid e\rangle _1\mid g\rangle _2+\beta
\mid e\rangle _1\mid s\rangle _2+\gamma \mid g\rangle _1\mid g\rangle_2
+\delta \mid g\rangle _1\mid s\rangle _2 \; ,
\end{equation}
where $\alpha$, $\beta$, $\gamma$ and $\delta$ are distributed according to the Haar
measure of $U(4)$ (that is, as resulting from the application of a random
Haar-distributed unitary on any fixed normalised state).
Let us recall that the Haar measure is defined as the measure which is
invariant under both left and right multiplication by any unitary transformation.
A distribution borrowed from the Haar measure is a natural choice for pure quantum states if one
does not want to privilege any specific direction in the Hilbert space,
and hence for testing the average fidelity of a given quantum operation,
as is the case here.
The desired, ideal output state $\ket{\Psi}_{out}$ of the {\sc CZ} gate, corresponding to
the input state $\ket{\Psi}_{in}$, is
\begin{equation}
\mid \Psi \rangle _{out}=\alpha \mid e\rangle _1\mid g\rangle _2-\beta
\mid e\rangle _1\mid s\rangle _2+\gamma \mid g\rangle _1\mid g\rangle_2
+\delta \mid g\rangle _1\mid s\rangle _2 \; ,
\end{equation}
where the phase flips only for the state $\mid e\rangle _1\mid s\rangle _2$.
\end{widetext}
Clearly, of the four superposed states defining $\ket{\Psi}_{in}$, only
$\ket{e}_1 \ket{g}_2$ and $\ket{e}_1\ket{s}_2$ evolve,
which further simplifies our task (besides the fact that only the single excitation subspace is
involved).
It is therefore straightforward to integrate Eq.~(\ref{master}) obtaining the final state $\varrho(t)$
of the system given the initial state
$\ket{\Psi}_{in}\ket{0}_{f}$, and to evaluate the Haar average of the fidelity:
\begin{equation}
\bar{F} = \int_{Haar} \bra{\Psi}_{out}
{\,\rm Tr}_{f}\left[\varrho(t)\right] \ket{\Psi}_{out} \,{\rm d}\ket{\Psi}_{in} \; ,
\end{equation}
where the notation $\int_{Haar}\,{\rm d}\ket{\Psi}_{in}$ loosely refers to the fact that integration is carried
over Haar-distributed input states, and ${\,\rm Tr}_{f}$ stands for the trace over the field's degrees of freedom.
In practice, this integral has been carried out by sampling the input states according to the Haar distribution
and by averaging the resulting fidelity. Samples of $200$ points turned out to give reliable
estimates, in that the average fidelity at the first two significant figures
would not change by increasing the sample's size.
Henceforth, we will always report fidelities at two decimal figures.
For some finer comparisons, we improved our sensitivity by increasing the
integration samples. In these cases as well
we will limit ourselves to two decimal digits, and will just qualitatively
point out the configurations providing higher fidelities.

\section{Results: Controlled coherent evolutions}

We will study now the approximated realization of the {\sc CZ} gate as detailed in the previous section, in several different situations, mainly varying the number $N$ of fibre modes and
their detuning with respect to the cavity modes.
All the results obtained are summarized at the end of the section.

Hereafter, we refer to the fibre
mode resonant with the atomic transition between $\mid e\rangle _k$ and $%
\mid g\rangle _k$ as the `central resonant mode'.
In order to find regimes with high average gate fidelity, we set $g_1= v _1=g$, while
changing $g_2$, $v_2$ and the interaction time $t$. To make comparisons more
clear, we define two parameters $\delta g=(g2-g)/g$ and $\delta
 v =( v _2-g)/g$.

\subsection{Two mediating modes. The Minimum Gap.}

To study coherent effects resulting from the competition of
multiple mediating modes, we start from the case of only two mediating
modes, with the same absolute detuning $\Delta$ with respect to
the frequency of the central resonant mode.
Fig.~\ref{fig3} shows the average gate fidelities $\bar{F}$
for different detunings at a time where the first peak in fidelity is achieved.
Clearly,
$\bar{F}$ decreases as the detuning $\Delta$ increases, since the two fibre
modes become more and more off-resonant.
For instance, at the peaks, $\bar{F}$ is $0.99$,
$0.98$ and $0.96$ for $\Delta/g =0.1,0.2,0.3$, respectively.
To put these fidelities into some context, let us mention that,
depending on the noise and on the affordable computational overhead,
error thresholds between $10^{-4}$ and $10^{-2}$ are believed to be required
to achieve fault-tolerant quantum computation \cite{errors}.

Typically, when $\Delta $ reaches $0.45g$,
and the gap between the two modes becomes comparable to the interaction strength $g$,
the average gate fidelity $\bar{F}$ drops to $0.90$.
As a reference, we will say that $0.9 g$ is the ``minimum gap'',
within which ``high fidelity'' ($90\%$) can still be recovered with two mediating off-resonant modes.
In a cavity QED implementation with $g \simeq 1$ GHz,
this gap would correspond to a fibre length $l$ of around one meter, since the spacing
of two neighbouring mediating modes is approximately $c\pi/l\simeq10^{9} {\,\rm Hz\, m} /l$.

\begin{figure}[t!]
\includegraphics[width=9cm,height=3cm]{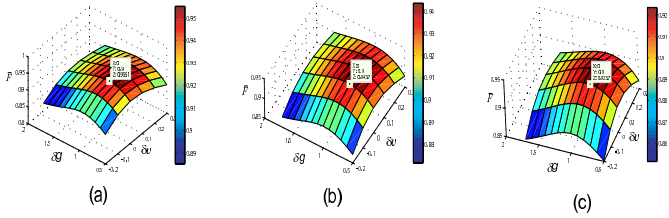}
\caption{(color online). The average gate fidelity $\overline{F}$ for two detuned modes versus $\delta g$ and $\delta v$  for loss rates $\gamma=\kappa=\Gamma=10^{-2}g$, and
for $\Delta=0.1g$ and $gt=4.6$ (a), $\Delta=0.2g$ and $gt=4.56$ (b),
and $\Delta=0.3g$ and $gt=4.54$ (c).} \label{fig4}
\end{figure}

As shown in Fig.~\ref{fig4},
$\bar{F}$ decreases
by around $4\%$, when losses are included for $\gamma
=\kappa =\Gamma =10^{-2}g$.
Further numerical analysis showed that for $\gamma
=\kappa =\Gamma =10^{-3}g$, the optimal fidelities decrease by around $0.5\%$.
These findings are in agreement with previous studies \cite{serafozzi} arguing that,
for nearly resonant couplings like the present one, average fidelities are virtually
unaffected for loss to coupling strengths ratios around $10^{-4}$,
are only slightly affected for ratios around $10^{-3}$, and start to affect the fidelity significantly -- between $1$ and $10\%$ -- when such ratios reach $10^{-2}$.
Of course, observing nearly ideal levels of coherence requires a high
degree of isolation. However, notice that a quantum {\sc CZ} gate implemented with
a fidelity of $85\%$ would still be a remarkable signature of quantum coherence,
implying the creation of substantial entanglement and of coherent off-diagonal
elements in the density matrix of the two qubits.

A note about the degree of stability of the gates obtained is also in order here:
a variation of the order of $0.1$ in the parameters
$\delta g$ and $\delta v $, corresponding to a variation of about $10\%$
in the coupling strengths, leads to a decrease in the gate fidelity of around $1\%$.
The same degree of sensitivity occurs
with respect to fluctuations in the interaction times.
Notice that, while achieving a lower peak in fidelity, the performance of more
off-resonant modes ($\Delta=0.3g$ in the figures) is slightly less sensitive to imperfections,
as evident from Fig.~\ref{fig3}.

\begin{figure}
\includegraphics[width=9cm]{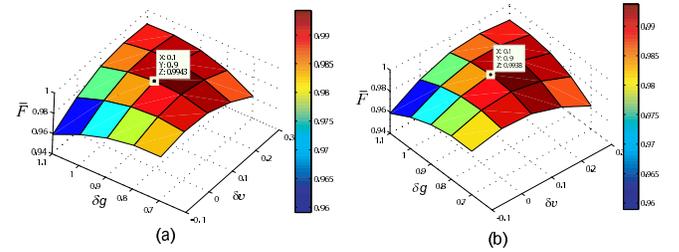}
\caption{(color online). The average gate fidelity $\bar{F}$
for a set of equally spaced mediating modes within the minimum gap $\Delta=0.45g$
versus $\delta g$ and $\delta v$ at the time $4.3 g^{-1}$, without considering losses.
In (a) 31 modes are considered; in (b) the central resonant mode has been removed.}\label{fig5}
\end{figure}

\subsection{Increasing number of mediating modes.}

\begin{figure}[b!]
\includegraphics[width=9cm]{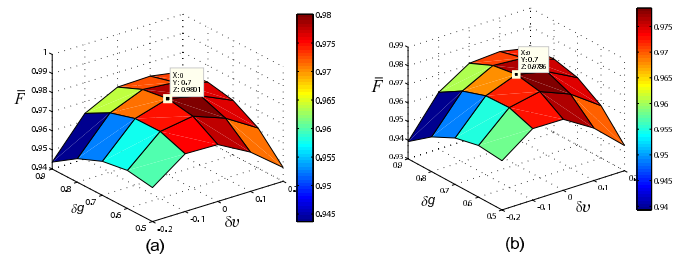}
\caption{(color online). The average gate fidelity $\bar{F}$ for equally spaced modes
outside the minimum gap, with detuning ranging
from $0.45g$ to $0.9g$, and from $-0.45g$ to $-0.9g$ versus $\delta g$, at the time $4.48 g^{-1}$,
without considering losses. In (b), 32 modes are considered;
in (a), a $33^{th}$ mode has been added, at the central resonant frequency.} \label{fig6}
\end{figure}

Let us now consider the case of a larger number of mediating modes.
Fig.~\ref{fig5} (a) shows the case of $31$ mediating modes,
with frequencies equally and symmetrically spaced around the central resonant mode
and filling the minimum gap.
At the peak, a remarkable average fidelity $\bar{F}=0.99$ can still be achieved:
the competition between the distinct modes, which could favour the `dispersion' of the
coherence, is clearly canceled out at optimal
times for such a configurations of closely packed, nearly resonant modes.
In the cavity QED model, for $g\simeq1{\,\rm GHz}$, the spacing considered here ($0.03g$)
would correspond to a fibre of approximately $30 {\,\rm m}$: in principle,
very long resonators can still mediate quantum evolutions coherently.
In Fig.~\ref{fig5} (b) the same case without the central resonant mode is depicted:
at the peak, the average fidelity drops very slightly but is still well above
$0.99$, which proves that the
coherent evolution mediated by these $31$ modes is not an effect due to the presence
of the central resonant mode.
Nor is this high fidelity a consequence of the symmetric distribution of the modes around the
central resonant frequency: this has been directly tested by shifting all the
frequencies of the mediating modes and does not produce any significant alteration in the
maximal average fidelity, as apparent from Fig.~\ref{fig9}.

To evaluate the influence of the modes very close to resonance with the cavity frequencies,
we now turn to cases where the set of mediating modes are all outside the `minimum gap'.
In Fig.\ref{fig6} (b), $32$ equally spaced modes
are considered: half of them spans the range between $0.45 g$ and $0.9 g$,
while the other half spans the range between $-0.45g$ and $-0.9g$.
Even in such an off-resonant configuration, a maximum average
fidelity of $0.98$ can be achieved.
This is a non-trivial finding, mostly if compared to the -- much lower -- fidelity
achievable with only two modes with frequencies {\em at the} minimum gap
(which is about $0.9$). In this case, more modes, and farther off from resonance,
allow one to achieve a better performance in terms of mediated quantum coherence.
This is therefore a remarkable instance where the cooperation between the mediating
modes prevails over their competition, and their effect could hence be harnessed
to implement distributed coherent quantum evolutions.
Fig.~\ref{fig6} (a) confirms that the addition of the central resonant mode has only
a relatively modest impact on the maximal fidelity (which raises slightly but is still essentially $0.98$).

\begin{figure}
\includegraphics[width=9cm]{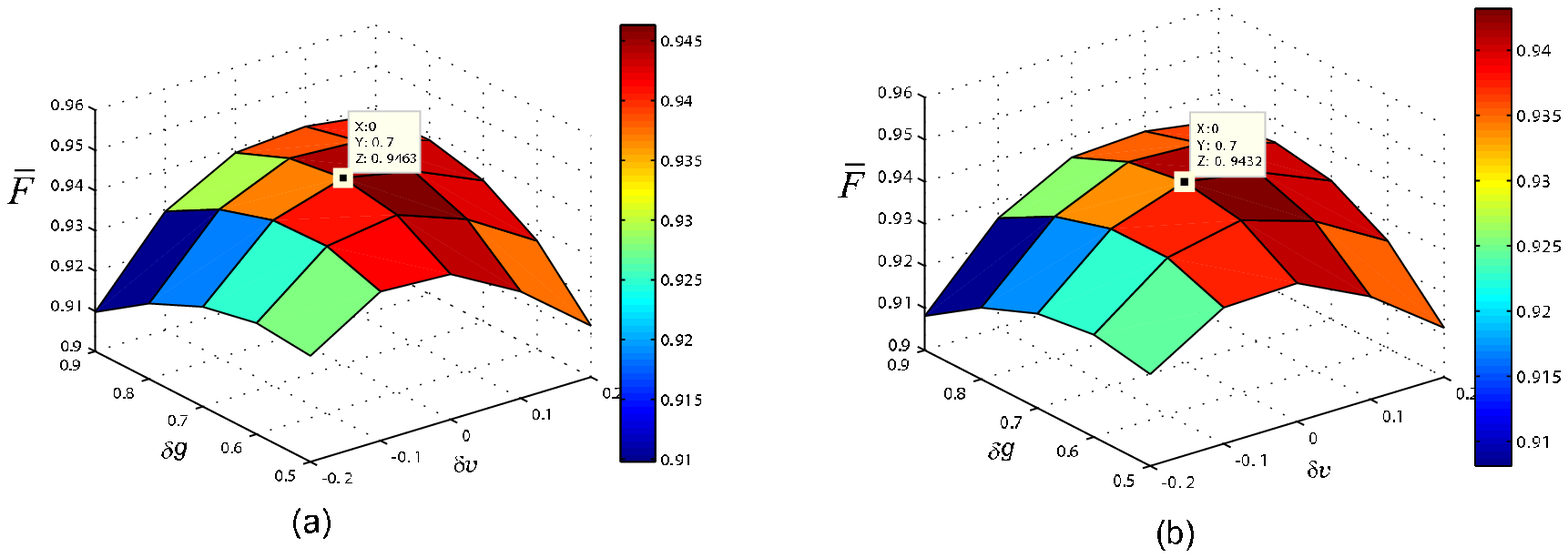}
\caption{(color online). The average gate fidelity $\bar{F}$ for
equally spaced modes outside the minimum gap, with detuning raging from
$0.45g$ to $0.9g$, and from $-0.45g$ to $-0.9g$, at the time $4.48 g^{-1}$, considering loss rates $\gamma=\kappa=\Gamma=10^{-2}g$. In (b), 32 modes are considered;
in (a), a $33^{th}$ mode has been added, at the central resonant frequency.}\label{fig7}
\end{figure}

Fig.~\ref{fig7} displays the effect of losses on the
optimal average fidelity:
for decay rates $\gamma = \kappa = \Gamma = 10^{-2}g$, the average gate fidelity $\bar{F}$
drops from around $0.98$ to around $0.94$.
Concerning decoherence, the resilience of many modes seems to be comparable to that of
few mediating modes.

\begin{figure}
\includegraphics[width=9cm]{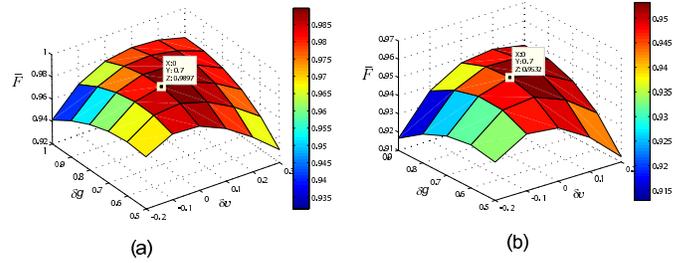}
\caption{(color online). The average gate fidelity $\bar{F}$
for 16 equally spaced mediating modes outside the minimum gap, with detuning ranging
from $0.45g$ to $0.9g$, and from $-0.45g$ to $-0.9g$ at the time $4.48 g^{-1}$.
No losses are considered in (a), while loss rates $\gamma=\kappa=\Gamma=10^{-2}g$
are introduced in (b).} \label{fig8}
\end{figure}

\begin{figure}[b!]
\includegraphics[width=9cm,height=5cm]{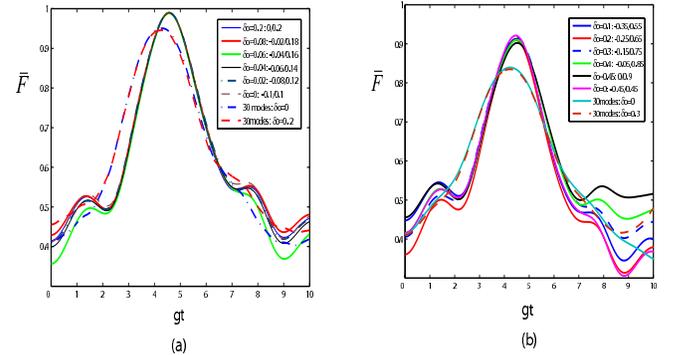}
\caption{(color online). The average gate fidelity $\overline{F}$ versus time for  $\delta g$ $=0.9$ and $\delta v=0$,
when the frequency spacing is kept constant but the centre of the set of frequencies is shifted.
In (a), the frequency difference between neighbouring modes is $0.2 g$ and the central frequency is shifted
from $0$ to $0.2 g$; the curves at the highest peak refer to two mediating modes,
while the two curves at the lowest peak refer to $30$ mediating modes.
In (b), the frequency difference between neighbouring modes is $0.9 g$ and the central frequency is shifted
from $0$ to $0.9 g$; the curves at the highest peak refer to two modes, while the two curves at the
lowest peak refer to $30$ modes. Notice that, mostly around the peaks, the curves are essentially
indistinguishable from each other.}\label{fig9}
\end{figure}

Finally, in Fig.~\ref{fig8}, the case of 16 modes with detuning ranging from $0.45g$
to $0.9g$ is considered. The results are very similar to the cases of 32 and 33
symmetrically detuned modes (including center resonant mode), in the same regime and
interaction time. At the peak, the average gate fidelity is 0.99.
Such a fidelity is reduced to $0.95$ when losses with $\gamma=\kappa=\Gamma=10^{-2}g$
are taken into account [see Fig.~\ref{fig8} (b)].

The situations addressed above, where finite regions of frequencies
are not populated, although of clear theoretical interest in the context of our study,
might seem rather artificial in practice.
However, even for bosonic fields, such situations
could be of relevance in systems like photonic crystals, where photonic bandgaps
arise for properly modulated refractive indexes \cite{bandgap}.

\subsection{`Many' mediating modes.}

\begin{figure}
\includegraphics[width=9cm]{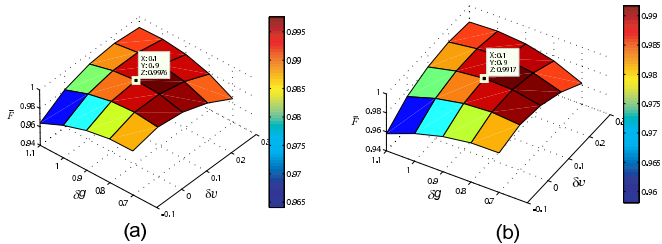}
\caption{(color online). The average gate fidelity $\bar{F}$
for a set of 100 equally spaced mediating modes
versus $\delta g$ and $\delta v$ at the time $4.3 g^{-1}$,
without considering losses.
In (a), the modes are
within the minimum gap, with detuning $\Delta$ from $-0.45g$ to $0.45g$.
In (b) the modes have detuning $\Delta$ ranging from $-1.5g$ to $1.5g$.
The plot (b) depicts the same case of Fig.~\ref{fig5}(b) but with the addition
of $70$ modes outside the minimum gap.}\label{fig10}
\end{figure}

Raising the number of mediating modes to $100$, one can see from Fig.\ref{fig10} (a) that
very high fidelities can be obtained, still above $0.99$, if all the
modes are taken within the minimum gap, which is promptly explained by the presence of
a large number of modes very close to resonance.
More interestingly, in Fig.~\ref{fig10} (b)
we have reconsidered the situation of Fig.~\ref{fig5} (b),
by adding to it $35$ modes in each direction with respect to the central resonant frequency.
These additional modes would of course be present in the realistic modelling of a
$300 {\, \rm m}$ long fibre, and could approximately account for the whole field
resonating in the fibre
(as the effect of more and more off-resonant modes can be considered to be very small).
In this case as well, the optimal average fidelity has been found to be well above $99\%$.

As for the stability of the quantum operations in the face of imperfections,
a comparative analysis of the plots shows that a larger number of modes
grant flatter and flatter fidelity peaks
[compare, in particular, the two peaks of Fig.~\ref{fig9}(a-b)].
So, while the fidelity obtained is slightly lower, the stability allowed by many mediating modes is
higher than that of two, or even one \cite{serafozzi}, mediating modes
(it can be inferred from Fig.~\ref{fig10} that, for $100$ modes,
a $10\%$ variation in the coupling strengths leaves the fidelity practically unchanged
at two decimal figures).

\subsection{Summary of results.}

Summing up, in this study we have shown that:
\begin{itemize}
\item{for only two mediating fibre modes, the optimal average fidelity $\bar{F}$ of a {\sc CZ}
decreases relatively rapidly with increasing gap between the modes, plummeting to $0.9$
for a gap equal to $0.9g$ (where $g$ is a reference value for the
atom-cavity and cavity-fibre couplings);}
\item{increasing the number $N$ of mediating modes, one can still achieve highly
reliable gates. For $N\simeq100$, $\bar{F}\simeq 0.99$.
This is the case regardless of the symmetry of the frequency spacing of
the mediating modes with respect to the central resonant mode, and of the presence of a
mode at resonance with the cavities;}
\item{nearly perfect gates ($\bar{F}\simeq0.98$ for $N\simeq30$)
can even be obtained for off-resonant interacting modes. In this case, a large number of modes
actually outperforms fewer modes;}
\item{loss rates around $10^{-3}g$ are necessary to operate such gates in
perfect conditions (to all practical purposes); however, even rates around $10^{-2}g$
allow for coherent effects to emerge (and entanglement to be generated);}
\item{for two mediating modes,
imperfections of the order of $10\%$ in the coupling strengths or interaction times
affect the resulting fidelity by approximately one percent; this stability improves for increasing number of modes, and the fidelity is virtually unchanged for imperfections around $10\%$ and $100$ mediating modes.}
\end{itemize}

\section{Conclusion and outlook}

By adopting a specific qubit encoding, and meaningfully restricting the system to the
single excitation subspace, we gave stringent evidence that distributed coherent evolutions
can be mediated by a large number of bosonic modes.

In practice this means that, if losses can be kept at bay, `very long' resonating structures
could be adopted to mediate distributed quantum computation
(or, conversely, that relatively low coupling strength could still be admissible).
In some particular configurations of the mediating frequencies, where a bandgap is present,
we also found cases where the collective effect of many modes outperforms
the mediation of fewer degrees of freedom,
thus dispelling the notion that employing fewer degrees of freedom
where, roughly speaking the `quantum information is less likely to disperse',
should result in better performances (it should be noted though, to avoid any ambiguity,
that all the fibre degrees of freedom {\em are} actually to be considered as `under control').

At a more fundamental level, this study shows that coherent quantum effect can be mediated
by many modes and, potentially, over long distances.

As a future step, we aim to extend this investigation beyond the restricted,
specific scheme we adopted here, to more general situations where distributed quantum
evolutions are mediated by a number $N$ bosonic modes.
We intend to address situations,
whose building block is essentially the ``spin-boson'' model \cite{spinboson},
where $N$ is large enough that it cannot be
treated with elementary techniques, but small enough not to allow a reliable description
of the field in terms of master equations for common baths,
like in \cite{braun,ucl}.
Let us mention that, very recently, the ground state properties of a ``two-impurity''
spin-boson model have been investigated by
adopting a variational approach \cite{ucl2}, and t-DMRG methods have been
adapted to the study of dimers strongly interacting with bosonic baths \cite{strongly}.
To tackle the study of the dynamics of such models, we will instead
resort to techniques based on co-moving bases and Ehrenfest dynamics \cite{dmitry}.
This investigation is currently under way \cite{future}.

\subsection*{Acknowledgments}

We thank A.~Bayat and H.~Wicherich for help with the numerics and
S.~Bose for a timely suggestion on how to speed them up.
SYY and AS also thank D.~Shalashilin, A.~Nazir and A.~Olaya-Castro for several fruitful
discussions on the general treatment of spin-boson-like systems, as well as E. Campbell and M. Hoban for valuable information about fault-tolerance thresholds.
SYY is supported by a KC Wong Scholarship.
ZBY and SBZ are supported by the National Natural Science
Foundation of China under grant no 10974028,
by the Fujian Natural Science Foundation of China under
grant no 2009J06002, and by the Doctoral Foundation of the Ministry
of Education of China under grants 20070386002 and
20093514110009. AS thanks the Central Research Fund
of the University of London for financial support.


\end{document}